\documentclass[10pt]{article}

\usepackage{color}
\usepackage{graphicx}
\usepackage{graphics}
\usepackage{amscd}
\usepackage{amsmath}
\usepackage{amsfonts}
\usepackage{amssymb}
\usepackage{cite}

\makeatletter
\newcommand*\bigcdot{\mathpalette\bigcdot@{.5}}
\newcommand*\bigcdot@[2]{\mathbin{\vcenter{\hbox{\scalebox{#2}{$\m@th#1\bullet$}}}}}
\makeatother

\DeclareMathOperator{\B}{\bf b}

\newcommand{\bF}{ {\mathbb F}}

\newcommand{\C}{ {\mathcal C}}
\newcommand{\EOP} { \hfill $\Box$ }

\newtheorem{theorem}{Theorem}[section]

\newtheorem{lemma}[theorem]{Lemma}

\newtheorem{proposition}[theorem]{Proposition}
\newtheorem{remark}[theorem]{Remark}

\begin{document}

\title{Rack-Aware MSR Codes with Multiple Erasure Tolerance}

\author{Jiaojiao Wang,\,\,\, Dabin Zheng{\thanks{Corresponding author. This work was partially supported by the National Natural Science Foundation of China under Grant Number 11971156.
\newline \indent ~~Jiaojiao Wang, Dabin Zheng and Shenghua Li are with the Hubei Key Laboratory of Applied Mathematics, Faculty of Mathematics and Statistics, Hubei University, Wuhan 430062, China (E-mail: wjiaojiao@stu.hubu.edu.cn, dzheng@hubu.edu.cn, lish@hubu.edu.cn )}}, \,\,\, Shenghua Li}

\date{ }

\maketitle

\begin{abstract}
The minimum storage rack-aware regenerating (MSRR) code is a variation of regenerating codes that achieves the optimal repair bandwidth for  a single node failure in the rack-aware model.
The authors in~\cite{Chen-Barg2019} and~\cite{Zhou-Zhang2021} provided explicit constructions of MSRR codes for all parameters to repair a single failed node. This paper generalizes
the results in~\cite{Chen-Barg2019} to the case of multiple node failures. We propose a class of MDS array codes and scalar Reed-Solomon (RS) codes, and show that these codes have optimal repair bandwidth
and error resilient capability for multiple node failures in the rack-aware storage model. Besides, our codes keep the same access level as the low-access constructions in~\cite{Chen-Barg2019} and~\cite{Zhou-Zhang2021}.
\end{abstract}



\section{Introduction}

Maximum distance separable (MDS) codes are widely used in modern large-scale distributed storage system for which provide the maximum failure tolerance for a given amount of storage overhead. An important metric of repair efficiency is the \emph{repair bandwidth}, i.e., the amount of data downloaded from other nodes for the purpose of the repair. Dimakis et al. in~\cite{Dimakis-Godfrey2010} have given a bound on the minimum number of symbols
required for repair of a single failed node, which is called the cut-set bound of the repair bandwidth. A repair scheme that attains this bound is called optimal and such codes are said to be \emph{minimum storage regenerating} (MSR) codes~\cite{Dimakis-Godfrey2010}. Over the last decade, important progresses have been made in the study of MSR codes, for example see \cite{Dimakis-Godfrey2010,Goparaju-Fazeli2017,Rashmi-Shah2011,Raviv-Silberstein2017,Tamo-Wang2013,Tamo-Ye2018,Ye-Barg2017,Ye-Barg20172} and reference therein. In addition, the basic repair problem of MDS codes has been extended to the case that some of the helper nodes provide erroneous information~\cite{Rashmi-Shah2012,Ye-Barg2017,Chen-Ye2020,Pawar-Rouayheb2011}.

In a homogeneous distributed storage model, all nodes  as well as communication between them are treated equally. However, modern data centers often have hierarchical
topologies by organizing nodes in racks, where the cross-rack communication cost is much more expensive than the intra-rack communication cost. This motives people to study
the repair problem for hierarchical data centers and many progresses have been made recently~\cite{Hou-Lee2019,Gupta-Lalitha2019,Hou-Lee2020,Hu-Lee2016,Jin-Luo2020,Chen-Barg2019,Pernas-Yuen2013,Sohn-Choi2018,Sohn-Choi2019,Tebbi-Chan2014,Zhang-Zhou2021,Zhou-Zhang2021}. In this paper, we focus on
the constructions and repair schemes of MSR codes in the rack-aware storage model.

Let $\C$ be an $(n, k, \ell)$ array code over a finite field $F$, i.e., a collection of codewords $c=(c_0, c_1, \cdots, c_{n-1})$, where each $c_i$ is a vector of length $\ell$ over $F$, or an element in the $\ell$ extension
of $F$. These code's coordinates are also called nodes. The amount of sub-packets stored in each node, i.e., $\ell$ is said to be \emph{sub-packetization}. A code $\C$ is called MDS if any $k$ coordinates of the codeword suffice to recover its remaining $n-k$ coordinates. Assume that $k$ data blocks are encoded into a codeword of length $n=u\bar{n}$ and stored across $n$ nodes. The $n$ nodes are organized equally into $\bar{n}$ groups, also called racks, and every rack contains $u$ nodes. The model of $u=1$ is the homogeneous case. To rule out the trivial case, we always assume that $u\leq k$, otherwise, a single node failure can be trivially repaired by $u-1$ surviving nodes within the same rack, and also assume that $u\leq n-k$ to ensure that code has repair ability if an entire rack fails. The rack which contains the failed nodes is called the \emph{host rack}.

The cut-set bound of repair bandwidth in a homogeneous distributed storage model has been generalized to the case of rack-aware storage model as follows~\cite{Hou-Lee2019,Chen-Barg2019}.
Let $k=\bar{k}u+v(0\leq v<u)$, where $u$ is the size of rack. Let $\beta_{u}(h,\bar{d})$ denote the minimum number of symbols over $F$ that one needs to download from the $\bar{d}$, $\bar{k}\leq\bar{d}\leq \bar{n}-1$ helper racks to recover $h$ failed nodes in the host rack. It was shown in~\cite{Chen-Barg2019} that
\begin{equation}\label{cut-set}
 \beta_{u}(h,\bar{d})\geq \frac{\bar{d}h\ell}{\bar{d}-\bar{k}+1}.
\end{equation}
A rack-aware repair scheme that achieves this bound is said to have optimal repair property and such codes are called \emph{minimum storage rack-aware regenerating}(MSRR) codes for repairing $h$ failed nodes in the rack-aware storage model.

In \cite{Ye-Barg2017}, Ye and Barg considered the error resilience capability in the repair process of multiple node failures in homogeneous distributed storage models. We generalize this concept to the case in rack-aware
storage models. Let $\bar{e}$ be a nonnegative integer with $0\leq\bar{e}\leq \lfloor \frac{\bar{d}-\bar{k}}{2}\rfloor$. Suppose that a subset of $\bar{e}$ racks out of $\bar{d}$ helper racks provide erroneous information and define $\beta_{u}(h,\bar{d},\bar{e})$ to be the minimum number of symbols needed from the helper racks to repair the $h$ failed nodes as long as the number of error racks in the helper racks is no more than $\bar{e}$.
For $\bar{k}+2\bar{e}\leq\bar{d}\leq \bar{n}-1$, the bound in (\ref{cut-set}) can be generalized to the following,
\begin{equation}\label{cut-setUER}
 \beta_{u}(h,\bar{d},\bar{e})\geq \frac{\bar{d}h\ell}{\bar{d}-2\bar{e}-\bar{k}+1}.
\end{equation}
An $(n, k, \ell)$ MDS code in rack-aware storage model is said to have \emph{universally error-resilient} (UER) $(h,\bar{d})$-optimal repair property if the equality in (\ref{cut-setUER}) holds.

Recently, Hou et al.~\cite{Hou-Lee2019} studied the parameters and constructions of MSRR codes, but their constructions need some constraints on the parameters and the finite fields being large enough.
The first explicit constructions of MSRR codes for all admissible parameters that support recovery of a single node failure were proposed by Chen et al. in~\cite{Chen-Barg2019}. Then Hou et al.~\cite{Hou-Lee2020} presented a coding framework that transformed an MSR code to an MSRR code. However, an MSRR code from such construction exists only if the finite field is sufficiently large.  Zhou et al.~\cite{Zhou-Zhang2021} provided another class of MDS array codes for all parameters in the rack-aware model with smaller sub-packetization and size of underlying finite field. As far as we know, it remains an open problem to construct MSRR codes with error resilience capability for supporting recovery of multiple node failures. In this paper, we propose a class of MDS array codes and RS codes in the rack-aware storage model and corresponding repair schemes such that the codes have UER $(h,\bar{d})$-optimal repair property when the number of failed nodes $h\leq u-v$, where $k=\bar{k}u+v(0\leq v<u)$. Moreover, we also provide repair schemes of discussed codes for the case $h> u-v$, and our schemes have the asymptotical UER $(h,\bar{d}+1)$-optimal repair property. Comparisons between our MSRR array code and previous constructions are shown in Table 1.

\begin{table}[!htb]
\centering
\caption{\small Comparisons with known MSRR codes, where $\bar{s}=\bar{d}-2\bar{e}-\bar{k}+1, k=\bar{k}u+v$, $h$ is the largest error tolerance.}
$\vspace{1mm}$
\begin{tabular}{|c|c|c|c|c|c|c|}
   \hline
   & sub-packet. $\ell$ & $\bar{d}$ & access per rack & $|F|$ & $h$ & UER \\
  \hline
  ~\cite{Chen-Barg2019} & $\bar{s}^{\bar{n}}$ & $ \bar{k}\leq\bar{d}\leq \bar{n}-1$ & $\frac{u\ell}{\bar{s}}$ & $u|(|F|-1)$,$|F|\geq n+\bar{s}-1$ & 1& -\\
  \hline
  ~\cite{Hou-Lee2020}& $\bar{s}^{\lceil\frac{\bar{n}}{\bar{s}}\rceil}$ & $\bar{d}=\bar{n}-1$ & $\frac{\ell}{\bar{s}}+(u-1)\ell$ & $|F|>k\ell\sum_{i=1}^{min\{k,\bar{n}\}}\binom{n-\bar{n}}{k-i}\binom{\bar{n}}{i}$ & 1 &- \\
  \hline
  ~\cite{Zhou-Zhang2021} & $\bar{s}^{\lceil\frac{\bar{n}}{u-v}\rceil}$ & $ \bar{k}\leq\bar{d}\leq \bar{n}-1$ & $\frac{u\ell}{\bar{s}}$ & $u|(|F|-1)$,$|F|> n$ & 1 &- \\
  \hline
  this paper & $\bar{s}^{\bar{n}}$ & $ \bar{k}\leq\bar{d}\leq \bar{n}-1$ & $\frac{u\ell}{\bar{s}}$ & $u|(|F|-1)$,$|F|> n$ & $u-v$ & \checkmark\\
  \hline
\end{tabular}
\end{table}
\normalsize

The rest of this paper is organized as follows. Section~\ref{Sec1} proposes a class of MDS array codes and  shows that they have the UER $(h,\bar{d})$-optimal repair property when the number of failed nodes $h\leq u-v$, and also
discusses the repair problem of the codes  when $h>u-v$ in the rack-aware storage system. In Section~\ref{Sec2} we show that RS codes in~\cite{Chen-Barg2019} have the UER $(h,\bar{d})$-optimal repair property in the rack-aware storage system. Section~\ref{sec:concluding} concludes this paper.

\section{MSRR codes with multiple erasure tolerance}\label{Sec1}

Let $F$ be a finite field of size $|F|> n$ and $u$ denote the number of nodes in each rack satisfying $u|(|F|-1)$. Let $n=u\bar{n}$ and $k=u\bar{k}+v$ for some $v$ with $0\leq v<u $. Let $r=n-k$ and $\bar{r}=\bar{n}-\bar{k}$.
 Let $\bar{d}$ denote the number of helper racks and $\bar{e}$ be the largest acceptable number of erroneous racks in the $\bar{d}$ helper racks satisfying $(u,\bar{s})=1$, where $\bar{s}=\bar{d}-2\bar{e}-\bar{k}+1$.
Let $\xi$ be a primitive element of $F$ and $\gamma$ be an element of $F$ with multiplicative order $u$. Denote $\lambda_{i,0}=\xi^i$ and $\lambda_{i,j}=1$ for $i\in \{0, 1,\cdots, \bar{n}-1\}$ and $j\in \{ 1, 2, \cdots, \bar{s}-1\}$. In this section, we propose a class of MDS array codes which are slightly modified codes in~\cite{Ye-Barg2017} and show that the codes have the UER ($h, \bar{d}$)-optimal repair property in rack-aware storage
 model for $h\leq u-v$. The objective array code is defined as follows.
\begin{equation}\label{eq:defcode}
 \C=\left\{ (C_{0},C_{1},\cdots,C_{n-1}): \sum_{i=0}^{\bar{n}-1}\sum_{g=0}^{u-1}A_{i,g}^{t}C_{iu+g}=0, \,\, t=0, 1, \cdots, r-1 \right\},
\end{equation}
here, $C_{j}$ is a column vector of length  $\ell=\bar{s}^{\bar{n}}$ over $F$, which denotes the $j$-th node and
\begin{equation}\label{eq:Aig}
A_{i,g}= \gamma^g A_i, \,\, \, A_i=\sum_{j=0}^{\ell-1}\lambda_{i,j_{i}}e_{j}e_{j(i,j_{i}\oplus 1)}^{T},\,\, i=0, 1, \cdots, \bar{n}-1, \,\, g=0,1,\cdots, u-1,
\end{equation}
where $\{e_{j}\, :\, j=0,1,\cdots,\ell-1\}$ is the standard basis of $F^{\ell}$ over $F$, $\oplus$ denotes addition modulo $\bar{s}$, $j_{i}$ is the $i$-th term of the base $\bar{s}$ expansion of $j=(j_{\bar{n}-1},\cdots, j_{1},j_{0})$ and $j(i, b)=(j_{\bar{n}-1},\cdots,j_{i+1}, b, j_{i-1},\cdots,j_{0})$, $b=0,1,\cdots,\bar{s}-1$.

By the similar discussion in Theorem VII.4 in \cite{Ye-Barg2017} we have the following proposition.
\begin{proposition}
The array code $\C$ given in (\ref{eq:defcode}) and (\ref{eq:Aig}) satisfies the MDS property.
\end{proposition}


To show the optimal repair property of the code $\C$, we need the following result.

\begin{lemma}(\cite{Ye-Barg2017})\label{bvandmatrix}  For two integers $n,\ell>0$, let $M_{0}, M_1, \cdots, M_{n-1}$ be $\ell$ order square matrices.
For any $i,j\in \{ 0, 1, \cdots, n-1\}$,  $M_{i}M_{j}=M_{j}M_{i}$ and $M_{i}-M_{j}$ is invertible, where $i\neq j$. Then
\begin{equation*}
  M=\left( \begin{array}{ccc}
                I_{\ell} & \cdots & I_{\ell} \\
                M_{0} & \cdots & M_{n-1} \\
                & \vdots &  \\
                M_{0}^{n-1} & \cdots &M_{n-1}^{n-1}
              \end{array}
  \right)
\end{equation*}
is invertible.
\end{lemma}

The following theorem is our main result in this section.

\begin{theorem}\label{thm:uer}
If the number $h$ of failed nodes located in the same rack satisfies $0<h\leq u-v$, then the array code $\C$ defined in (\ref{eq:defcode}) and (\ref{eq:Aig}) has the UER $ (h,\bar{d})$-optimal repair property.
\end{theorem}
{\it Proof.}  First we repair the linear combination of the nodes in the host rack from the remaining $\bar{n}-1$ surviving racks. Denote $\Lambda_{i,j_{i}, 0}=1$ and $\Lambda_{i,j_{i},t}=\lambda_{i,j_{i}}\lambda_{i,j_{i}\oplus1}\cdots\lambda_{i,j_{i}\oplus(t-1)}$ for $t=1, 2, \cdots, r-1$. By direct calculations, we  have
\begin{equation*}
   A_{i,g}^{t}=\left( \sum_{j=0}^{\ell-1}\lambda_{i,j_{i}}\gamma^{g}e_{j}e_{j(i,j_{i}\oplus 1)}^{T}\right)^{t}=\sum_{j=0}^{\ell-1}\Lambda_{i,j_{i},t}\gamma^{gt}e_{j}e_{j(i,j_{i}\oplus t)}^{T}.
\end{equation*}
 The $(iu+g)$-th node $C_{iu+g}$ can be rewritten as $\sum_{j=0}^{\ell-1}c_{iu+g,j}e_{j}$. By combining these with (\ref{eq:defcode}), we get
 \begin{equation}\label{eq:defc1}
\sum_{j=0}^{\ell -1} \sum_{i=0}^{\bar{n}-1}\sum_{g=0}^{u-1}\gamma^{gt} \Lambda_{i,j_{i},t} c_{iu+g,j(i,j_{i}\oplus t)}e_{j}=0,\quad t=0,1,\cdots,r-1.
\end{equation}
 Without loss of generality, suppose that the $(\bar{n}-1)$-th rack is the host rack. The equality (\ref{eq:defc1}) is rewritten coordinate wise as the following,
\begin{equation}\label{eq:defcoordinatewise}
\Lambda_{\bar{n}-1,j_{\bar{n}-1},t}\sum_{g=0}^{u-1}\gamma^{gt} c_{(\bar{n}-1)u+g,j(\bar{n}-1,j_{\bar{n}-1}\oplus t)}
=-\sum_{i=0}^{\bar{n}-2}\Lambda_{i,j_{i},t}\sum_{g=0}^{u-1}\gamma^{gt}c_{iu+g,j(i,j_{i}\oplus t)},
\end{equation}
where $t=0, 1, \cdots, r-1$ and $j=0, 1, \cdots,\ell-1$. Consider the parity-check equations in (\ref{eq:defcoordinatewise}) for $t\in \{ m, u+m,\cdots,(\bar{s}-1)u+m\}$, where $m$ is a fixed number in $\{0,1,\cdots,u-v-1\}$.
Since $\gamma^{u}=1$, from (\ref{eq:defcoordinatewise}) we have
\begin{equation}\label{eq:defcoordinatewise1}
\Lambda_{\bar{n}-1,j_{\bar{n}-1},wu+m}\sum_{g=0}^{u-1}\gamma^{gm}c_{(\bar{n}-1)u+g,j(\bar{n}-1,j_{\bar{n}-1}\oplus(wu+m))} =-\sum_{i=0}^{\bar{n}-2}\Lambda_{i,j_{i},wu+m}\sum_{g=0}^{u-1}
\gamma^{gm}c_{iu+g,j(i,j_{i}\oplus (wu+m))},
\end{equation}
where  $w=0, 1, \cdots, \bar{s}-1$ and $j=0, 1, \cdots,\ell-1$. So, $\sum_{g=0}^{u-1}\gamma^{gm}c_{(\bar{n}-1)u+g,j(\bar{n}-1, j_{\bar{n}-1}\oplus(wu+m))}$ can be determined by $\{\sum_{g=0}^{u-1}\gamma^{gm}c_{iu+g,j(i,j_{i}\oplus(wu+m))}\,\, :\,\, i=0,1,\cdots,\bar{n}-2\}$. Since $(u,\bar{s})=1$ and $m<u$, for any fixed $j_i$ and $m$, we have that
$\{ j_i \oplus (wu+m)\, :\, w=0,1, \cdots, \bar{s}-1\} = \{0, 1, \cdots, \bar{s}-1 \}$. Set $\ell^\prime =\bar{s}^{\bar{n}-1}$ and $j_{\bar{n}-1}=0$ in (\ref{eq:defcoordinatewise1}), then we have that $\{\sum_{g=0}^{u-1}\gamma^{gm}c_{(\bar{n}-1)u+g,j}:j=0,1,\cdots,\ell-1\}$ can be derived from $\{\sum_{g=0}^{u-1}\gamma^{gm}c_{iu+g,j}\, :\, i=0,1,\cdots,\bar{n}-2; j=0,1,\cdots,  \ell^\prime-1\}$,
that is to say, the linear combination of nodes in the $(\bar{n}-1)$-th rack can be determined by the vectors of length $\ell^\prime$ in the first $\bar{n}-1$ racks.


Then we prove that any $\bar{d}$ ($\bar{k}\leq \bar{d}\leq \bar{n}-1$) helper racks out of $\bar{n}-1$ surviving racks can recover the values $\sum_{g=0}^{u-1}\gamma^{gm}c_{(\bar{n}-1)u+g,j}, \,j=0,1,\cdots,\ell-1$.
To this end, set $C_{iu+g}^{(\ell^\prime)}=(c_{iu+g,0},c_{iu+g,1},\cdots,c_{iu+g,\ell^\prime-1})^{T}$, where $i=0, 1, \cdots, \bar{n}-2$ and define
\begin{equation}\label{eq:defcm}
\C_{m}=\left( \sum_{g=0}^{u-1}\gamma^{gm}C_{g}^{(\ell^\prime)}, \sum_{g=0}^{u-1}\gamma^{gm}C_{u+g}^{(\ell^\prime)}, \cdots,\sum_{g=0}^{u-1}\gamma^{gm}C_{(\bar{n}-2)u+g}^{(\ell^\prime)} \right),
\,\, m=0, 1, \cdots, u-v-1.
\end{equation}
Assume that there are $\bar{d}$ helper racks and at most $\bar{e}$ out of $\bar{d}$ helper racks with $\bar{k}+2\bar{e} \leq \bar{d} \leq \bar{n}-1$ have errors.
Next, we show that $\C_m$ is an $(\bar{n}-1,\bar{d}-2\overline{e},\ell^\prime)$ MDS array code for any fixed $m\in \{0, 1, \cdots, u-v-1\}$, that is to say, any $\bar{d}-2\bar{e}$ columns
can represent all columns in $\C_m$.

Consider the parity-check equations of $\C$ in~(\ref{eq:defcode}) for $t \in\{ m, u+m,\cdots,u(\bar{r}-\bar{s}-1)+m\}$, then
\begin{equation}\label{con221}
  \sum_{i=0}^{\bar{n}-1}A_{i}^{u\eta+m}\sum_{g=0}^{u-1}\gamma^{gm}C_{iu+g}=0,\,\, \eta=0,1,\cdots,\bar{r}-\bar{s}-1.
\end{equation}
Since
$$A_{i}^{u\bar{s}}=(\sum_{j=0}^{\ell-1}\lambda_{i,j_{i}}e_{j}e_{j(i,j_{i}\oplus 1)}^{T})^{u\bar{s}}=\sum_{j=0}^{\ell-1}(\Lambda_{i,j_{i},\bar{s}})^{u}e_{j}e_{j}^{T}=\xi^{iu}I_{\ell},$$
from the parity-check equations of $\C$ in~(\ref{eq:defcode}) for $t \in \{ u\bar{s}+m, u(\bar{s}+1)+m,\cdots,u(\bar{r}-1)+m\}$, we have
\begin{equation}\label{con222}
\sum_{i=0}^{\bar{n}-1}A_{i}^{u(\eta+\bar{s})+m}\sum_{g=0}^{u-1}\gamma^{gm}C_{iu+g}=\sum_{i=0}^{\bar{n}-1}\xi^{iu}A_{i}^{u\eta+m}\sum_{g=0}^{u-1}\gamma^{gm}C_{iu+g}=0,\,\, \eta=0,1,\cdots,\bar{r}-\bar{s}-1.
\end{equation}
Multiplying $\xi^{(\bar{n}-1)u}$ on the both sides of (\ref{con221}) and then subtracting (\ref{con222}) we get
\begin{equation}\label{con223}
 \sum_{i=0}^{\bar{n}-2}(\xi^{(\bar{n}-1)u}-\xi^{iu})A_{i}^{u\eta+m}\sum_{g=0}^{u-1}\gamma^{gm}C_{iu+g}=0.
\end{equation}
Substituting $A_i$ in (\ref{eq:Aig}) into (\ref{con223}), we have
\begin{equation}\label{con224}
\begin{split}
    0 &= \sum_{i=0}^{\bar{n}-2}\left(\xi^{(\bar{n}-1)u}-\xi^{iu}\right) \left(\sum_{j=0}^{\ell-1}\lambda_{i,j_{i}}e_{j}e_{j(i,j_{i}\oplus1)}^{T}\right)^{u\eta+m}\sum_{g=0}^{u-1}\gamma^{gm}\left(\sum_{j=0}^{\ell-1}c_{iu+g,j}e_{j}\right) \\
      &= \sum_{i=0}^{\bar{n}-2}\left(\xi^{(\bar{n}-1)u}-\xi^{iu}\right)\sum_{g=0}^{u-1}\gamma^{gm}\left( \sum_{j=0}^{\ell-1}\Lambda_{i,j_{i},u\eta+m}c_{iu+g,j(i,j_{i}\oplus(u\eta+m))}e_{j}\right).
\end{split}
\end{equation}
The equality (\ref{con224}) is rewritten coordinate wise as the following:
\begin{equation}\label{con225}
\sum_{i=0}^{\bar{n}-2}(\xi^{(\bar{n}-1)u}-\xi^{iu})\sum_{g=0}^{u-1}\gamma^{gm}\Lambda_{i,j_{i},u\eta+m}c_{iu+g,j(i,j_{i}\oplus(u\eta+m))}=0,\,\, j=0,1,\cdots,\ell-1.
\end{equation}
Let
\begin{equation}\label{eq:Bi}
B_{i}=\sum_{j=0}^{\ell^\prime -1}\lambda_{i,j_{i}}e_{j}^{(\ell^\prime)}(e_{j(i,j_{i}\oplus 1)}^{(\ell^\prime)})^{T},\,\, i=0,1,\cdots,\bar{n}-2,
\end{equation}
where $\{e_{j}^{(\ell^\prime)}\, :\, j=0,1,\cdots,\ell^\prime-1\}$ is a set of standard basis column vectors in $F^{\ell^\prime}$. It is known that $B_{i}$ is the leading principle submatrix of $A_{i}$
with order $\ell^\prime$. By direct verifications, from (\ref{con225}) we have
\begin{equation}\label{Bi parity check}
\begin{split}
&\sum_{i=0}^{\bar{n}-2}\left(\xi^{(\bar{n}-1)u}-\xi^{iu}\right)B_{i}^{u\eta+m}\sum_{g=0}^{u-1}\gamma^{gm}C_{iu+g}^{(\ell^\prime)}\\
=&\sum_{i=0}^{\bar{n}-2}\left(\xi^{(\bar{n}-1)u}-\xi^{iu}\right) \sum_{g=0}^{u-1}\gamma^{gm}\left(\sum_{j=0}^{\ell^\prime-1}\lambda_{i,j_{i}}e_{j}^{(\ell^\prime)}(e_{j(i,j_{i}\oplus 1)}^{(\ell^\prime)})^{T}\right)^{u\eta+m}\sum_{j=0}^{\ell^\prime-1}c_{iu+g,j}e_{j}^{(\ell^\prime)}\\
=&\sum_{i=0}^{\bar{n}-2}(\xi^{(\bar{n}-1)u}-\xi^{iu})\sum_{g=0}^{u-1}\gamma^{gm}\sum_{j=0}^{\ell^\prime-1}\Lambda_{i,j_{i},u\eta+m}c_{iu+g,j(i,j_{i}\oplus(u\eta+m))}e_{j}^{(\ell^\prime)}\\
=&0,
\end{split}
\end{equation}
where $\eta=0,1,\cdots,\bar{r}-\bar{s}-1$.

Choose $\bar{d}-2\bar{e}$ columns in $\C_m$ with index set $H=\{p_{0}, p_1, \cdots,p_{\bar{d}-2\bar{e}-1}\}$. Let $\{ q_{0}, q_1, \cdots, q_{\bar{r}-\bar{s}-1}\} = \{0, 1, \cdots, \bar{n}-2\} \setminus H$.
Then~(\ref{Bi parity check}) can be rewritten as follows:
\begin{equation*}
\sum_{i=0}^{\bar{r}-\bar{s}-1}(\xi^{(\bar{n}-1)u}-\xi^{q_{i}u})B_{q_{i}}^{u\eta+m}\sum_{g=0}^{u-1}\gamma^{gm}C_{q_{i}u+g}^{(\ell^\prime)}=
-\sum_{j=0}^{\bar{d}-2\bar{e}-1}(\xi^{(\bar{n}-1)u}-\xi^{p_{j}u})B_{p_{j}}^{u\eta+m}\sum_{g=0}^{u-1}\gamma^{gm}C_{p_{j}u+g}^{(\ell^\prime)}.
\end{equation*}
These $\bar{r}-\bar{s}$ equations are rewritten as the matrix equation form as follows:
\begin{equation}\label{Bicoef}
 \begin{split}
 \underbrace{\begin{pmatrix}
I_{\ell^\prime}&\cdots&I_{\ell^\prime}
 \\ B_{q_{0}}^{u}&\cdots&B_{q_{\bar{r}-\bar{s}-1}}^{u}\\\vdots&&\vdots
 \\B_{q_{0}}^{(\bar{r}-\bar{s}-1)u}&\cdots&B_{q_{\bar{r}-\bar{s}-1}}^{(\bar{r}-\bar{s}-1)u}\end{pmatrix}}_B  \times
 \underbrace{\begin{pmatrix}(\xi^{(\bar{n}-1)u}-\xi^{q_{0}u})B_{q_{0}}^{m}&&&\\&\ddots&&\\&& \ddots &\\&& (\xi^{(\bar{n}-1)u}-\xi^{q_{\bar{r}-\bar{s}-1}u})B_{q_{\bar{r}-\bar{s}-1}}^{m} \end{pmatrix}}_D \\
 \times\begin{pmatrix}\sum_{g=0}^{u-1}\gamma^{gm}C_{q_{0}u+g}^{(\ell^\prime)}\\\vdots\\ \vdots\\\sum_{g=0}^{u-1}\gamma^{gm}C_{q_{\bar{r}-\bar{s}-1}u+g}^{(\ell^\prime)}\end{pmatrix}
 =-\begin{pmatrix}
  \sum_{j=0}^{\bar{d}-2\bar{e}-1}(\xi^{(\bar{n}-1)u}-\xi^{p_{j}u})B_{p_{j}}^{m}\sum_{g=0}^{u-1}\gamma^{gm}C_{p_{j}u+g}^{(\ell^\prime)}
  \\\vdots\\ \vdots \\\sum_{j=0}^{\bar{d}-2\bar{e}-1}(\xi^{(\bar{n}-1)u}-\xi^{p_{j}u})B_{p_{j}}^{u(\bar{r}-\bar{s}-1)+m}\sum_{g=0}^{u-1}\gamma^{gm}C_{p_{j}u+g}^{(\ell^\prime)}\end{pmatrix}.
\end{split}
\end{equation}
 By the definition of $B_i$ in (\ref{eq:Bi}), it is easy to verify that $B_i$ and $ B_i^u -B_j^u$ are invertible, and $B_i^u B_j^u = B_j^u B_i^u$
for $j\neq i$. So, the matrix $B$ is invertible by Lemma~\ref{bvandmatrix}, and then the matrix $B\times D$ is invertible. Therefore, the columns in $\C_m$ with index set $H$ can represent
all columns in $\C_m$, that is to say, $\C_m$ is an $(\bar{n}-1,\bar{d}-2\bar{e},\ell^{\prime})$ MDS array code.

Choosing any $\bar{d}$ columns from $\C_m$ also constitutes a $(\bar{d},\bar{d}-2\bar{e},\ell^\prime)$ MDS array code, which is a punctured code of $\C_m$. This code is viewed as a linear code
over the $\ell^\prime$ extension of $F$. Then its minimum distance is $2\bar{e}+1$, and so can correct $\bar{e}$ errors. Therefore, by downloading any $\bar{d}$ nodes in $\C_m$,
we can recover the entire codeword as long as the number of erroneous nodes among the $\bar{d}$ helper nodes is not greater than $\bar{e}$, and
further recover the linear combination of nodes in the $(\bar{n}-1)$-th rack $\{\sum_{g=0}^{u-1}\gamma^{gm}c_{(\bar{n}-1)u+g,j}\,:\,j=0,\cdots,\ell-1\}$ from (\ref{eq:defcoordinatewise1}).

Finally, we recover the $h$ specific failed nodes in the host rack. 
Assume that the index set of failed nodes in the host rack (i.e., the $(\bar{n}-1)$-th rack) is $\mathcal{F}=\{g_1, g_2, \cdots, g_h\}$ and $\mathcal{T}=\{ 0, 1, \cdots, u-1\} \setminus \mathcal{F}$.
We have that $\Delta_m =\sum_{g=0}^{u-1}\gamma^{gm}c_{(\bar{n}-1)u+g,j}$ is known for $j=0,\cdots,\ell-1$ and $m =0,1, \cdots, u-v-1$. Taking $m=0, 1, \cdots, h-1$ we get
\begin{equation}\label{eq:linearsysh}
\begin{gathered}
 \begin{pmatrix}1&\cdots&1
 \\\gamma^{g_{1}}&\cdots&\gamma^{g_{h}}\\\vdots&&\vdots
 \\\gamma^{g_{1}(h-1)}&\cdots&\gamma^{g_{h}(h-1)}
 \end{pmatrix}
 \begin{pmatrix}c_{(\bar{n}-1)u+g_{1},j}\\
 c_{(\bar{n}-1)u+g_{2},j}\\
 \vdots \\c_{(\bar{n}-1)u+g_{h},j}
 \end{pmatrix}
 =\begin{pmatrix}\Delta_0- \sum_{g\in \mathcal{T}}c_{(\bar{n}-1)u+g,j}\\
 \Delta_1- \sum_{g\in \mathcal{T}}\gamma^g c_{(\bar{n}-1)u+g,j}\\
 \vdots\\
 \Delta_{h-1}- \sum_{g\in \mathcal{T}} \gamma^{g(h-1)}c_{(\bar{n}-1)u+g,j}\end{pmatrix}, \,\, j=0,1, \cdots, \ell-1 .
 \end{gathered}
\end{equation}
Since $\gamma^{g_{i}}\neq\gamma^{g_{i'}}$ for all $i, i'\in \{ 0,1, \cdots, u-1\}$ with $i\neq i'$, from the linear system (\ref{eq:linearsysh}), we can recover the failed nodes in the host rack with index set $\mathcal{F}$.

To recover theses nodes we have downloaded $\bar{d} h\ell^\prime=\frac{\bar{d}\ell h}{\bar{s}}$ symbols. This value is exactly the lower bound in~(\ref{cut-setUER}), and so the discussed code has the UER $(h,\bar{d})$-optimal repair property. \EOP


\begin{remark} In the repair process described above, to recover the $h$, $0<h\leq u-v$ failed nodes of the code $\C$ defined in (\ref{eq:defcode}) and (\ref{eq:Aig}), we have accessed the symbols in the set $\{c_{iu+g}:i\in \mathcal{R}; g=0,\cdots,u-1;j=0,\cdots,\ell'-1\}$, where $\mathcal{R}$ is the index set of $\bar{d}$ helper racks. Then the total number of the accessed symbols is
$\frac{\bar{d}u\ell}{\bar{s}}$. So, this code has the same low-access property as that of the codes constructed in~\cite{Chen-Barg2019} and~\cite{Zhou-Zhang2021}.
\end{remark}

The discussion above shows that the MDS array code $\C$ defined in (\ref{eq:defcode}) and (\ref{eq:Aig}) has optimal repair property when the number of failed nodes in the host rack is no more than $u-v$.
When the number of failed nodes $h$ is greater than $u-v$, the code $\C$ has asymptotical UER ($h,\bar{d}+1$)-optimal repair property, i.e., when the number of helper racks $\bar{d}+1$ is large enough, the ratio between the amount of download symbols and the optimal bound in~(\ref{cut-setUER}) approaches 1.


\begin{theorem}\label{thm con2}
If the number $h$ of failed nodes located in the same rack satisfies $u-v<h\leq u$, then the repair bandwidth of the array code $\C$ defined in (\ref{eq:defcode}) and (\ref{eq:Aig}) is less than $\frac{(\bar{d}+1)\ell h}{\bar{s}}$ and the code $\C$ has an error correction capability.
\end{theorem}
{\it Proof.} The repair scheme consists of the following two main steps.

(1) For $m=0, 1, \cdots, u-v-1$, by a similar calculation in Theorem~\ref{thm:uer} we can recover $\sum_{g=0}^{u-1}\gamma^{gm}c_{(\bar{n}-1)u+g,j}$, $j=0,1,\cdots, \ell-1$ from any $\bar{d}$ out of $\bar{n}-1$ columns in $\mathcal{C}_{m}$ defined in~(\ref{eq:defcm}) as long as the number of helper racks where errors occur is no more than $\bar{e}$.

(2) For $m=u-v, u-v+1, \cdots, h-1$, the values of $\eta$ in~(\ref{con221}) and~(\ref{con222}) could be $0, 1,\cdots,\bar{r}-\bar{s}-2$.
We can get $\sum_{g=0}^{u-1}\gamma^{gm}c_{(\bar{n}-1)u+g,j}$, $j=0,1,\cdots, \ell-1$ from any $\bar{d}+1$ out of $\bar{n}-1$ columns in $\mathcal{C}_{m}$ as long as the number of helper racks where errors occur is no more than $\bar{e}$. Repeating the calculations in~(\ref{con223})-(\ref{Bicoef}), we can show that $\mathcal{C}_{m}$ defined in~(\ref{eq:defcm}) for $m\in \{ u-v, u-v+1, \cdots, h-1\}$ is also an $(\bar{n}-1,\bar{d}+1-2\bar{e},\ell^\prime)$ MDS array code. So, we can get $\sum_{g=0}^{u-1}\gamma^{gm}c_{(\bar{n}-1)u+g,j}$, $j=0,1,\cdots, \ell-1$ from any $\bar{d}+1$ out of $\bar{n}-1$ columns in $\mathcal{C}_{m}$. By a similar argument to that in Theorem~\ref{thm:uer}, we know that
this code can correct at most $\bar{e}$ errors among $\bar{d}+1$ helper racks. Then, by the similar calculation in equation~(\ref{eq:linearsysh}), we can recover the $h$ failed nodes.

In the two steps above, we download $\frac{\bar{d}\ell(u-v)}{\bar{s}}$ and $\frac{(\bar{d}+1)\ell(h-u+v)}{\bar{s}}$ symbols from $\bar{d}$ and $\bar{d}+1$ helper racks, respectively.
So, to recover the $h$ failed nodes we have altogether downloaded $\frac{\bar{d}\ell(u-v)}{\bar{s}}+\frac{(\bar{d}+1)\ell(h-u+v)}{\bar{s}}=\frac{\bar{d}\ell h}{\bar{s}}+\frac{\ell(h-u+v)}{\bar{s}}$ symbols. Note that $v<u$, then $h-u+v<h$. Therefore, $\frac{\bar{d}\ell h}{\bar{s}}+\frac{\ell(h-u+v)}{\bar{s}}<\frac{(\bar{d}+1)\ell h}{\bar{s}}$. In this case, the ratio of the amount of download symbols to the optimal repair bandwidth
given in (\ref{cut-setUER}) is less than $1+\frac{1}{\bar{d}-2\bar{e}-\bar{k}+1}$. So, the repair bandwidth of the code approaches optimal level if the number of helper racks is large enough when $h> u-v$.
\EOP

\section{Rack-aware RS codes with multiple erasure tolerance}\label{Sec2}

Reed-Solomon codes, the most practically used MDS codes, have been employed in many distributed storage systems. Guruswami and Wootters first proposed the optimal repair scheme of RS codes
for the homogeneous distributed storage system~\cite{Guruswami-Wootters2017}. Chen and Barg in \cite{Chen-Barg2019} studied the repair procedure of RS codes having optimal repair property
for a single node failure in the rack-aware storage system. This section generalizes the repair process of RS codes in~\cite{Chen-Barg2019} to the case of the multiple node failures. We propose a new repair scheme for the RS codes defined in~\cite{Chen-Barg2019} which can optimally repair multiple failed nodes from arbitrary $\bar{d}$ helper racks. Moreover, the error correction capability and the low-access property of these RS codes are discussed.

Let $q$ be a power of a prime and $\bF_q$ be a finite field of $q$ elements. Let $u$ be the size of the rack with $u|(q-1)$ and $n = \bar{n}u$. Let $k=\bar{k}u+v$, $0\leq v<u$. Let $\bar{d}$ denote the number of helper racks and $\bar{e}(0\leq\bar{e}\leq \lfloor \frac{\bar{d}-2\bar{k}}{2}\rfloor )$ be the largest acceptable number of erroneous racks in the $\bar{d}$ helper racks. Let $\bar{s}=\bar{d}-2\bar{e}-\bar{k}+1$ and $p_{0},\cdots,p_{\bar{n}-1}$ be $\bar{n}$ distinct primes such that
$$p_{i}\equiv 1\,\, {\rm mod}\,\, \bar{s}\,\, {\rm and} \,\, p_{i}>u , \,\, i=0,1,\cdots,\bar{n}-1. $$
Let $\gamma$ be an element in $\bF_{q}$ with the multiplicative order $u$ and $\lambda_{i}$ an element of degree $p_{i}$ over $\bF_{q}$ for $i=0,1, \cdots, \bar{n}-1$. Let
\[ F_{i}:=\bF_{q}(\lambda_{j}\,:\, j\in\{ 0,1,\cdots,\bar{n}-1\}\setminus\{i\}), \,\, i=0,1,\cdots,\bar{n}-1 \,\,  {\rm and } \,\, F:=\bF_{q}(\lambda_{0},\lambda_{1},\cdots,\lambda_{\bar{n}-1}). \]
Let $K$ be an extension of $F$ of degree $\bar{s}$ and $\alpha\in K$ be a generating element of $K$ over $F$, i.e., $K=F(\alpha)$. Thus, for any $i\in \{0,1, \cdots, \bar{n}-1\}$ we have the chain of
inclusions
\[ \bF_q \subset F_i \subset F \subset K .\]
So, $K$ is the $\ell$-th degree extension of $\bF_q$, where $\ell = \bar{s}\prod_{i=0}^{\bar{n}-1} p_i$ is called sub-packetization. Choosing the set of evaluation points as $\Omega=\{\lambda_{0}\gamma^{0},\cdots,\lambda_{0}\gamma^{u-1},\cdots,\lambda_{\bar{n}-1}\gamma^{0},\cdots,\lambda_{\bar{n}-1}\gamma^{u-1}\}$, we define the objective RS code as follows:
\begin{equation}\label{eq:rscode}
\C =\left\{ \left( f(\lambda_{0}\gamma^{0}),\cdots, f(\lambda_{0}\gamma^{u-1}), \cdots, f(\lambda_{\bar{n}-1}\gamma^{0}), \cdots, f(\lambda_{\bar{n}-1}\gamma^{u-1})\right)\,:\, f(x)\in K[x], {\rm deg}(f)<k \right\}.
\end{equation}

To describe the repair procedure for multiple {node failures} in rack-aware model, we first recall some necessary preliminaries.

\begin{lemma}[\cite{Tamo-Ye2018}]\label{lem4}
For $i\in \{ 0,1,\cdots,\bar{n}-1\}$, consider the $F_{i}$-linear subspace $S_{i}$,
\begin{equation*}\label{S}
S_{i}={\rm Span}_{F_{i}}\left\{ \sum_{m=0}^{\bar{s}-1}\alpha^{m}\lambda_{i}^{u(p_{i}-1)}, \alpha^{j}\lambda_{i}^{u(j+t\bar{s})},\, \, j=0,1,\cdots,\bar{s}-1,\,\, t=0,1,\cdots,\frac{p_{i}-1}{\bar{s}}-1\right\}.
\end{equation*}
Then
$$ {\rm dim}_{F_i} S_i = p_i, \,\, S_{i}+S_{i}\lambda_{i}^{u}+\cdots+S_i \lambda_{i}^{u(\bar{s}-1)}=K,$$
where $S_{i}\beta =\{\theta \beta, \theta \in S_{i}\}$, and the operation $+$ is the Minkowski sum of sets, $T_{1}+T_{2}=\{t_{1}+t_{2}:t_{1}\in T_{1},t\in T_{2}\}$.
\end{lemma}

\begin{lemma}[\cite{Lidl-Niederreiter1996}]\label{lem3}
Let $\mathbb{F}$ be a finite field and $\mathbb{K}$ an $m$-degree extension of $\mathbb{F}$. Let $\{\xi_{1}, \xi_2, \cdots,\xi_{m}\}$ be a basis of $\mathbb{K}$ over $\mathbb{F}$, and $\{\bar{\xi}_{1}, \bar{\xi}_2, \cdots, \bar{\xi}_{m}\}$ its dual basis. Then for any $\beta\in \mathbb{K}$, we have
$$\beta =\sum_{t=1}^{m}tr_{\mathbb{K}/\mathbb{F}}\left( \xi_{t}\beta \right)\bar{\xi}_{t}.$$
\end{lemma}

\begin{theorem}\label{thm:RS}
If the number $h$ of failed nodes located in the same rack satisfies $0<h\leq u-v$, then the RS code $\C$ defined in (\ref{eq:rscode}) has the UER $(h,\bar{d})$-optimal repair property. The repair procedure accesses $\ell/\bar{s}$ symbols on each of the nodes in the $\bar{d}$ helper racks, and the repair scheme is independent of the choice of the subset of $\bar{d}$ helper racks.
\end{theorem}
{\it Proof.} The coordinates of every codeword in $\C$ are viewed as vectors over the field $\bF_{q}$, and they also represent data on each node. The node size equals $\ell=\bar{s}\prod_{i=0}^{\bar{n}-1}p_{i}$, which is the extension degree of $K$ over $\bF_{q}$.

Let $\C^{\perp}$ denote the dual code of $\C$.
It is known that $\C^{\perp}$ is a generalized RS code, which has the following form:
\begin{equation}\label{eq:rs dual}
\begin{split}
\C^{\perp}=&\Big\{ \left(v_{0,0} f(\lambda_{0}\gamma^{0}),\cdots, v_{0,u-1}f(\lambda_{0}\gamma^{u-1}), \cdots, v_{\bar{n}-1, 0}f(\lambda_{\bar{n}-1}\gamma^{0}), \cdots, v_{\bar{n}-1, u-1} f(\lambda_{\bar{n}-1}\gamma^{u-1})\right)\, : \\
 &\,\,\, f(x)\in K[x], {\rm deg}(f)<n-k \Big\},
\end{split}
\end{equation}
where $v_{i,j}$ are nonzero in $K$ and can be represented by $\lambda_i$ and $\gamma^j$ for $i\in \{0,1, \cdots, \bar{n}-1\}$ and $j\in \{ 0, 1, \cdots, u-1\}$.

Denote $r=n-k$. Take $f(x)$ in~(\ref{eq:rs dual}) to  be $x^t$, where $t=0,1,\cdots,r-1$.  For every codeword $(C_{0}, C_{1},\cdots, C_{n-1})\in \C$, we have
\begin{equation}\label{eq:orth1}
\sum_{i=0}^{\bar{n}-1}\sum_{g=0}^{u-1}v_{i,g}(\lambda_{i}\gamma^{g})^{t}C_{iu+g}=0,\,\,\,  t=0,1,\cdots, r-1.
\end{equation}
Assume that the $i^*$-th rack is the host rack.  Let $\{ e_{0}, e_{1},\cdots, e_{p_{i^{*}}-1}\}$ be a basis of the vector space $S_{i^{*}}$ over the field $F_{i^{*}}$. From (\ref{eq:orth1}) we have
\begin{equation}\label{eq:orth2}
e_{j}\sum_{i=0}^{\bar{n}-1}\sum_{g=0}^{u-1}v_{i,g}(\lambda_{i}\gamma^{g})^{t}C_{iu+g}=0,\,\,t=0,1,\cdots, r-1,
\end{equation}
where $j\in \{ 0,1,\cdots,p_{i^{*}}-1\}$. Consider the subset of the parity-check equations in (\ref{eq:orth2}) with indices $t=m,u+m,\cdots,(\bar{r}-1)u+m$ for some fixed $m\in \{0,1,\cdots,u-v-1\}$, then
\begin{equation}\label{eq:orth3}
e_{j}\lambda_{i^{*}}^{uw+m}\sum_{g=0}^{u-1}v_{i^{*},g}\gamma^{gm}C_{i^{*}u+g}= -e_{j}\sum_{i\neq i^{*}} \lambda_{i}^{uw+m}\sum_{g=0}^{u-1}v_{i,g}\gamma^{gm}C_{iu+g}, \,\, w=0,1,\cdots,\bar{r}-1.
\end{equation}
 Let $tr_{i^{*}}(\cdot)=tr_{K/F_{i^{*}}}(\cdot)$ be the trace mapping from $K$ to $F_{i^{*}}$. Since $\lambda_{i}\in F_{i^{*}}$ for $i\neq i^{*}$ and $\gamma\in \bF_{q}$, applying $tr_{i^{*}}(\cdot)$ to the both sides of (\ref{eq:orth3}), we have
\begin{equation}\label{eq:orthtrace1}
tr_{i^{*}}\left( e_{j}\lambda_{i^{*}}^{uw+m}\sum_{g=0}^{u-1}v_{i^{\ast},g}\gamma^{gm}C_{i^{*}u+g}\right)
=-\sum_{i\neq i^{*}}\lambda_{i}^{uw+m}\sum_{g=0}^{u-1}\gamma^{gm}tr_{i^{*}}\left( e_{j}v_{i,g}C_{iu+g}\right), \,\, w=0,1,\cdots,\bar{r}-1,
\end{equation}
where $j\in \{ 0,1,\cdots,p_{i^{*}}-1\}$. For a given $m\in \{0,1,\cdots,u-v-1\}$, from~(\ref{eq:orthtrace1}) we can recover the set $\{tr_{i^{*}}(e_{j}\lambda_{i^{*}}^{uw+m}\sum_{g=0}^{u-1}v_{i^{\ast},g}\gamma^{gm}C_{i^{*}u+g}):w=0,1,\cdots,\bar{r}-1\}$
from $\{\sum_{g=0}^{u-1}\gamma^{gm}tr_{i^{*}}(e_{j}v_{i,g}C_{iu+g}):i\in [0,\bar{n}-1]\setminus \{i^{*}\}\}$, where $j\in \{ 0,1,\cdots,p_{i^{*}}-1\}$. Since $\bar{s}=\bar{d}-2\bar{e}-\bar{k}+1\leq\bar{r}$, we can recover the set
\begin{equation}\label{eq:sequations}
  \left\{ tr_{i^{*}}(e_{j}\lambda_{i^{*}}^{uw+m}\sum_{g=0}^{u-1}v_{i^{\ast},g}\gamma^{gm}C_{i^{*}u+g})\,:\, w=0,1,\cdots,\bar{s}-1, \,\,  j=0, 1, \cdots,p_{i^{*}}-1 \right\}
\end{equation}
from the set
\begin{equation}\label{eq:barneqs}
\left\{ \sum_{g=0}^{u-1}\gamma^{gm}tr_{i^{*}}(e_{j}v_{i,g}C_{iu+g})\,:\, i\in \{ 0, 1, \cdots, \bar{n}-1\} \setminus \{i^{*}\}, \,\, j=0, 1, \cdots,p_{i^{*}}-1 \right\}.
\end{equation}
By Lemma~\ref{lem4}, we know that $\{e_{j}\lambda_{i^{*}}^{uw+m}\,:\,j=0, 1, \cdots,p_{i^{*}}-1;w=0,\cdots,\bar{s}-1\}$ is a basis of $K$ over $F_{i^{*}}$. By choosing its dual basis,
from Lemma~\ref{lem3} we can recover $\sum_{g=0}^{u-1}v_{i^{\ast},g}\gamma^{gm}C_{i^{*}u+g}$ from the set in (\ref{eq:sequations}).
So, obtaining the values in the set in (\ref{eq:barneqs}) efficiently is the key to recovering the linear combination of the failed nodes in the host rack, i.e.,$\sum_{g=0}^{u-1}v_{i^{\ast},g}\gamma^{gm}C_{i^{*}u+g}$.

Next, we will discuss the process to repair the set in (\ref{eq:barneqs}) with the minimum amount of download symbols in the case of errors in helper racks. For all $i\in \{ 0, 1, \cdots,\bar{n}-1\} \setminus \{i^*\}$ and  $m\in \{ 0,1,\cdots,u-v-1\}$, we define $u-v$ array codes as follows:
\begin{equation}\label{eq:rsmds}
  \C_{m}=\left( \Upsilon_{m,0},\cdots,\Upsilon_{m,i^{*}-1},\Upsilon_{m,i^{*}+1},\cdots,\Upsilon_{m,\bar{n}-1} \right),\,\, m=0,1,\cdots,u-v-1,
\end{equation}
where $\Upsilon_{m,i}=(\sum_{g=0}^{u-1}\gamma^{gm}tr_{i^{*}}(e_{j}v_{i,g}C_{iu+g}),j=0,1,\cdots,p_{i^{*}}-1)^{\perp}$. By slight modifying the proof of the case in homogeneous storage case in ~\cite[Sec.\uppercase\expandafter{\romannumeral3}]{Chen-Ye2020} we can show that $\C_{m}$ in (\ref{eq:rsmds}) is an $(\bar{n}-1,\bar{d}-2\bar{e},p_{i^{*}})$ MDS array code for any $m\in\{0, 1, \cdots,u-v-1\}$.
So, for an integer $\bar{d}$ with $\bar{k}+2\bar{e}\leq \bar{d}\leq \bar{n}-1$, any $\bar{d}$ out of $\bar{n}-1$ columns in $\C_m$ suffice to recover all columns of $\C_m$ as long as the number of errors in the $\bar{d}$ columns is not greater than $\bar{e}$. Hence, in such case, $\sum_{g=0}^{u-1}v_{i^{*},g}\gamma^{gm}C_{i^{*}u+g}$ can be recovered from $\bar{d}$ columns in $\C_m$ for any $m\in \{ 0,1,\cdots,u-v-1\}$.
By similar calculations to that in Theorem~\ref{thm:uer}, we can recover the $h$ failed nodes in the host rack from known linear combinations $\sum_{g=0}^{u-1}v_{i^{*},g}\gamma^{gm}C_{i^{*}u+g}, \, m=0, 1, \cdots, u-v-1$.
Moreover, to repair $h$ failed nodes in the same rack, we have downloaded $\frac{\bar{d}\ell h}{\bar{s}}$ symbols over $\bF_q$ from the helper racks, and this meets the cut-set bound in~(\ref{cut-setUER}). So, the RS code in (\ref{eq:rscode}) has the optimal repair bandwidth and error correction capability when the number of helper racks where the error occurred is no more than $\bar{e}$.

During the repair process described above, the amount of access symbols is $\bar{d} u \ell$. In the following we further discuss how to reduce this amount. Let $p*=\prod_{i=0}^{\bar{n}-1} p_i/p_{i^*}$ and let
$\{\varepsilon_0, \varepsilon_1, \cdots, \varepsilon_{p^*-1}\}$ be a basis of $F_{i^*}$ over $\bF_q$ and $\{\varepsilon_0^*, \varepsilon_1^*, \cdots, \varepsilon_{p^*-1}^*\}$ its dual basis. By Lemma~\ref{lem3},
\begin{equation}\label{eq:orthtracereps}
\begin{split}
\sum_{g=0}^{u-1}\gamma^{gm} tr_{i^{*}}(e_{j}v_{i,g}C_{iu+g}) &= \sum_{t=0}^{p^*-1}tr_{F_{i^{*}}/\bF_{q}}\left(\varepsilon_{t} ( \sum_{g=0}^{u-1}\gamma^{gm}tr_{i^{*}}(e_{j}v_{i,g}C_{iu+g}))\right) \varepsilon_{t}^{*} \\
       & =\sum_{g=0}^{u-1}\gamma^{gm}\sum_{t=0}^{p^*-1}tr_{K/\bF_{q}}\left( \varepsilon_{t}e_{j}v_{i,g}C_{iu+g}\right) \varepsilon_{t}^{*},
   \end{split}
\end{equation}
where $j=0,1,\cdots, p_{i^{*}}-1$. It is easy to show that the vectors $\varepsilon_t e_j, t=0, 1, \cdots, p^*-1, j =0, 1, \cdots, p_{i^*}-1$ are in $K$ and linearly independent over $\bF_q$. We expand these vectors to a basis of
$K$ over $\bF_q$, and denote it by $\{ \beta_i, i=0, 1, \cdots, \ell-1\}$, and its dual basis is represented as $\{ \beta_i^*, i=0, 1, \cdots, \ell-1\}$. For any $i\neq i^{*}$ and $g\in \{ 0, 1, \cdots, u-1\}$, the element $v_{i,g}C_{iu+g}$ can be represented as follows:
\begin{equation}\label{eq:spandC}
 v_{i,g}C_{iu+g}=\sum_{b=0}^{\ell-1}c_{iu+g,b}\beta_{b}^{*},
\end{equation}
where $c_{iu+g,b}\in \bF_{q}$. Substituting~(\ref{eq:spandC}) into~(\ref{eq:orthtracereps}) we have
\begin{equation}\label{eq:accessnodes}
\begin{split}
\sum_{g=0}^{u-1}\gamma^{gm} tr_{i^{*}}(e_{j}v_{i,g}C_{iu+g}) &= \sum_{g=0}^{u-1}\gamma^{gm}\sum_{t=0}^{p^*-1}tr_{K/\bF_{q}}\left( \varepsilon_{t}e_{j}\sum_{b=0}^{\ell-1}c_{iu+g,b}\beta_{b}^{*}\right)\varepsilon_{t}^{*} \\
& =\sum_{g=0}^{u-1}\gamma^{gm}\sum_{t=0}^{p^*-1}\sum_{b=0}^{\ell-1}tr_{K/\bF_{q}}(\varepsilon_{t}e_{j}\beta_{b}^{*})c_{iu+g,b} \varepsilon_{t}^{*}.
   \end{split}
\end{equation}
The equality (\ref{eq:accessnodes}) shows that $\sum_{g=0}^{u-1}v_{i^{\ast},g}\gamma^{gm}C_{i^{*}u+g}$ can be repaired by accessing the symbols $c_{iu+g,b}$ for which $tr_{K/\bF_{q}}(\varepsilon_{t}e_{j}\beta_{b}^{*})\neq 0$
for $t=0, 1, \cdots, p^*-1, j =0, 1, \cdots, p_{i^*}-1$ and $b=0, 1, \cdots, \ell-1$. Recall that all $\varepsilon_{t}e_{j}$ are in set $\{ \beta_i, i=0, 1, \cdots, \ell-1\}$ and its dual basis is $\{ \beta_b^*, b=0,1 \cdots, \ell-1\}$. So, the amount of access symbols is $\bar{d}u p^* p_{i^{*}}=\bar{d}u\ell/\bar{s}$. This amount of access symbols is the same as the low-access construction in~\cite{Chen-Barg2019} and~\cite{Zhou-Zhang2021}.
\EOP

When the number of failed nodes in host rack is more than $u-v$, the discussed code has asymptotical UER ($h,\bar{d}+1$)-optimal repair property. Combining the repair process in Theorem~\ref{thm con2} and Theorem~\ref{thm:RS}, we have the following result.

\begin{theorem}\label{RS TH2}
If the number $h$ of failed nodes located in the same rack satisfies $u-v<h\leq u-1$, then the repair bandwidth of the RS code $\C$ defined in (\ref{eq:rscode}) is less than $\frac{(\bar{d}+1)\ell h}{\bar{s}}$. Moreover, this code has the error correction capability and accesses $\ell/\bar{s}$ symbols on each of the nodes in $\bar{d}+1$ helper racks in the process of repair.
\end{theorem}

\section{Concluding remark }\label{sec:concluding}

In this paper we proposed a class of MDS array codes and RS codes in the rack-aware storage system, and showed that they have the UER $(h,\bar{d})$-optimal repair property when the number of failed nodes $h\leq u-v$. When $u-v<h\leq u$, the discussed codes have asymptotical UER $(h,\bar{d}+1)$-optimal repair property. It is worthy of further designing MSRR codes and the corresponding repair scheme such that they have smaller sub-packetization and optimal access property.

\end{document}